| | |
|---|---|
| Title | Etching Processes of Polytetrafluoroethylene Surfaces Exposed to He and He-O$_2$ Atmospheric Post-discharges |
| Authors | J. Hubert, T. Dufour, N. Vandencasteele, S. Desbief, R. Lazzaroni and F. Reniers |
| Affiliations | [1] Faculté des Sciences, Service de Chimie Analytique et Chimie des Interfaces, Université Libre de Bruxelles, CP-255, Bld du Triomphe 2, B-1050 Bruxelles, Belgium<br>[2] Service de Chimie des Matériaux Nouveaux, Université de Mons-UMONS/Materia Nova, 20 Place du Parc, 7000 Mons, Belgium |
| Ref. | Langmuir, 2012, Vol. 28, Issue 25, 9466-9474 |
| DOI | http://dx.doi.org/10.1021/la300822j |
| Abstract | A comparative study of polytetrafluoroethylene (PTFE) surfaces treated by the post-discharge of He and He-O$_2$ plasmas at atmospheric pressure is presented. The characterization of treated PTFE surfaces and the species involved in the surface modification are related. In pure He plasmas, no significant change of the surface has been observed by X-ray photoelectron spectroscopy (XPS), dynamic water contact angles (dWCA) and atomic force microscopy (AFM), in spite of important mass losses recorded. According to these observations, a layer-by-layer physical etching without any preferential orientation is proposed, where the highly energetic helium metastables are the main species responsible for the scission of –(CF$_2$)$_n$– chains. In He–O$_2$ plasmas, as the density of helium metastables decreases as a function of the oxygen flow rate, the treatment leads to fewer species ejected from the PTFE surfaces (in agreement with mass loss measurements and the detection of fluorinated species onto aluminum foil). However, the dWCA and AFM measurements show an increase in the hydrophobicity and the roughness of the surface. The observed alveolar structures are assumed to be caused by an anisotropic etching where the oxygen atoms etch mainly the amorphous phase. |

# 1. Introduction

For many years, scientists have used plasmas to modify polymer surfaces without affecting their bulk properties.[1,2] However, the phenomena occurring at the interface between the plasma and the polymer are still not well understood. In most cases, a polymer exposed to a reactive plasma shows a decrease in water contact angle (WCA), due to the grafting of selected chemical functions depending on the nature of the plasma gas.[1,3–5]

Many studies concerning the plasma surface modification of polytetrafluoroethylene (PTFE) have already been published.[6–18] Nevertheless, treatments of PTFE by oxygen containing plasmas are more controversial, as both hydrophobic and hydrophilic modifications are obtained. For instance, Ryan et al.[14] observed that an oxygen plasma treatment induces a roughening of the samples with no change in the chemical composition. This behavior was similar to those reported by Morra et al.[17] for long treatment times. On the other hand, other studies[12,15,16] showed that oxygen radio-frequency (RF) plasma could lead to an etching of the surface characterized by oxygen grafting and a decrease in WCA. Even if the mechanism remains mostly unknown, some researchers have suggested interesting theories. Recently, Vandencasteele et al.[7] highlighted in this journal the synergetic role of charged species (electrons) and atomic oxygen in the etching of PTFE by low-pressure oxygen plasmas. In a previous study,[6] we corroborated the necessity of oxygen to improve the hydrophobicity of PTFE, and we suggested a mechanism of anisotropic etching in an atmospheric pressure He−O$_2$ post-discharge.

Contrary to oxygen-containing plasmas, helium plasmas operated at atmospheric pressure usually lead to quite a strong decrease (50°) in WCA with an incorporation of oxygen.[19,20] As far as we know, the mechanism of modification has not been discussed further.

Nevertheless, helium has been identified to be responsible for a sputtering of PTFE in low-pressure plasma21,22 as PTFE-like films have been deposited.

In this paper, we study the influence of the addition of $O_2$ in a He plasma for the post-discharge treatment of PTFE. As almost no modification of dynamic water contact angles (dWCA), X-ray photoelectron spectroscopy (XPS), and atomic force microscopy (AFM) analyses are observed in the absence of oxygen, we expected no major influence of He plasmas on PTFE surfaces. However, the mass loss study and detection of ejected species by an Al trap presented in this paper shows a stronger sputtering in the case of a pure He plasma. We also identify by optical emission spectroscopy (OES), the species that could be involved in the surface modifications of PTFE with and without oxygen.

## 2. Experimental section

### 2.1. Samples

One mm thick PTFE samples were supplied by Goodfellow. The samples were cleaned in pure methanol (AnalaR Normapur, VWR) and pure iso-octane (GR for analysis, Merck), before being exposed to the plasma post-discharge.

### 2.2. Plasma source

The sample surfaces are exposed to the linear post-discharge of an RF atmospheric plasma torch, the Atomflo 400L-Series, from SurfX Technologies23 which has been detailed in a previous paper.6 Helium (vector gas) and oxygen (reactive gas) are studied for flow rates ranging from 10 to 20 L/min and from 0 to 200 mL/min, respectively. The RF power commonly used is comprised between 60 and 130 W. The geometry of the slit is described as "linear" due to the ratio of its aperture length (20 mm) to its width (0.8 mm). In order to avoid possible temperature effects, experiments were performed 20 min after the ignition of the plasma torch.

A robotic system is integrated to the plasma torch, enabling the treatment of large samples located downstream of the discharge. The scanning treatment is achieved with respect to 3 degrees of freedom corresponding to the three axes of a Cartesian coordinate system. In all our experiments, the plasma source was only moved along one direction. The robotic system allows one to vary the following parameters: the scanning velocity (vs), the scanning length (Ls), the number of scans (Ns), and the gap (gs) corresponding to the distance between the plasma torch and the upper surface of the PTFE sample. Although two different gaps (500 μm and 1 mm) have been used due to practical constraints (for instance, the OES optical fiber requires a minimum gap of 1 mm), we previously showed that the behaviors, regarding dWCA and XPS, were similar for these two gap values.

### 2.3. Diagnostics

A drop shape analyzer (Krüss DSA 100) was used to measure dWCAs on PTFE samples according to the dynamic sessile drop method. Advancing and receding contact angles were both measured by growing and shrinking the size of a single drop on the PTFE surface, from 0 to 15 μL and back to 0 μL at a rate of 30 μL/min. Advancing contact angles are the maximum angles observed during the droplet growth. Receding contact angles are considered as the minimum angle at the three-phases contact line during the drop withdrawal. In order to avoid wrong values coming from the deformation of the drop, all the receding contact angles have systematically been measured 20 s after the beginning of the shrinking of the drop (10 s before total absorption).

To evaluate the chemical composition at the surface of the samples, XPS analyses were performed on a Physical Electronics PHI-5600 instrument with a base pressure of $\sim 10^{-9}$ mbar in the analysis chamber. Survey scans were used to determine the chemical composition of the elements present at the PTFE surface. Narrow-region photoelectron spectra were used for the chemical study of the C 1s, O 1s, and F 1s peaks. Spectra were acquired using the Mg anode (1253.6 eV) operating at 300 W. Wide surveys were acquired at 93.9 eV pass-energy, with a five-scan accumulation (time/step: 50 ms, eV/step: 0.8), and high-resolution (HR) spectra of the C1s peaks at 23.5 eV pass-energy with an accumulation of 10 scans (time/step: 50 ms, eV/step: 0.025). The elemental composition was calculated after the removal of a Shirley background and using the sensitivity coefficients coming from the manufacturer's handbook: $S_C=0.205$, $S_F=1.000$, and $S_O=0.63$.

AFM was used to analyze the surface morphology of the PTFE samples. AFM images were recorded in air with a Nanoscope IIIa microscope operated in tapping mode. The probes were commercially available silicon tips with a spring constant of 24–52 N/m, a resonance frequency lying in the 264–339 kHz range, and a typical radius of curvature in the 5–10 nm range. The images presented here are height images recorded with a sampling resolution of 512 × 512 data points and a scan size of 5 × 5 µm$^2$.

The PTFE samples were weighted before and after the plasma treatments to evaluate mass variations. For this we used the Sartorius BA110S Basic series analytical balance characterized by a capacity of 110 g and a readability of 0.1 mg. Moreover, during the plasma treatment, aluminum foil was placed close to the samples. As aluminum is known to be an efficient fluorine trap,[24] we then analyzed the foil by XPS, looking for the presence of fluorinated species.

OES was performed with a SpectraPro-2500i spectrometer from ACTON research Corporation (0.500 m focal length, triple grating imaging). The light emitted by the post-discharge was collected by an optical fiber and transmitted to the entrance slit (50 µm) of the monochromator. There, the light is collimated, diffracted, focused on the exit slit, and finally captured by a charge-coupled device (CCD) camera from Princeton Instruments. Each optical emission spectrum was acquired with the 1800 grooves.mm$^{-1}$ grating (blazed at 500 nm) and recorded on 30 accumulations with an exposure time of 25 ms. As the recorded emission intensity is a function of the total post-discharge emission, which was different for every $O_2$ flow rate, the emissions of all the species were normalized to the emission of the whole post-discharge (i.e., a continuum ranging from 250 to 850 nm).

## 3. Results & Discussion

The results are presented in three sections by studying the surface modification of PTFE samples in direct and indirect ways. The first part of this work is dedicated to the direct modifications of PTFE surfaces by the post-discharge of an atmospheric plasma torch supplied in helium with or without oxygen. The second part discloses indirect measurements to characterize the surface modifications of PTFE. In the third part, we will focus on species involved at the plasma/PTFE interface.

## 3.1. Surface modifications of PTFE

In order to characterize the evolution of the hydrophobicity of the PTFE samples, dWCA measurements were carried out. Figure 1 shows the effect of the $O_2$ flow rate onto the surface modification of PTFE. A low hysteresis was observed between the advancing and receding dWCA. The treatments with high oxygen flow rates are characterized by high WCAs (around 140°), and a plateau is reached for $O_2$ flow rates above 100 mL/min. Moreover, smaller hysteresis are observed as the oxygen flow rate is increased. This trend could be explained by the higher (super)hydrophobic character of the surface, thus approaching the Cassie−Baxter model.[25] Indeed, we experimentally observed that the drop could be unstable and easily slid from the PTFE surface, which is characteristic of a non-wetted contact.

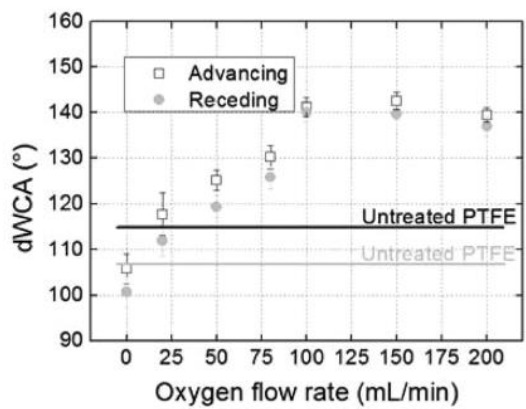

Figure 1. Influence of oxygen flow rate on the dWCA [$v_s$ = 25 mm/s, $L_s$ = 10 mm, $g_s$ = 500 μm, $N_s$ =1000, $Φ(He)$ = 15 L/min, $P_{RF}$ = 90W].

As previously shown,[6,7,10] the presence of oxygen is required to obtain more hydrophobic surfaces. In the case of a pure helium plasma (oxygen flow rate = 0 mL/min), a decrease in dWCA of almost 10° compared with the native PTFE dWCA (~115°) is observed. The general decrease in dWCA by pure helium plasma at atmospheric pressure is in good agreement with the literature,[19,20] even if in their case, much lower WCA values were reached (~50°).

To complete these results, XPS survey and HR spectra were recorded to determine the chemical composition of the PTFE surface. In our work, the surface composition and the surface chemistry (Figures 2 and 3) do not seem to be affected by the plasma conditions, as the F/C ratio stays constant and the spectral envelope of the HR C 1s spectra are not modified by the plasma exposure. On the contrary, in the literature,[19,20] the HR C1s XPS spectrum of samples treated by a pure helium plasma indicated the presence of oxygenated compounds. The observed differences from the literature could be explained by the use of a post-discharge in this research, which involves a softer treatment of the surface compared to the plasma conditions previously described.

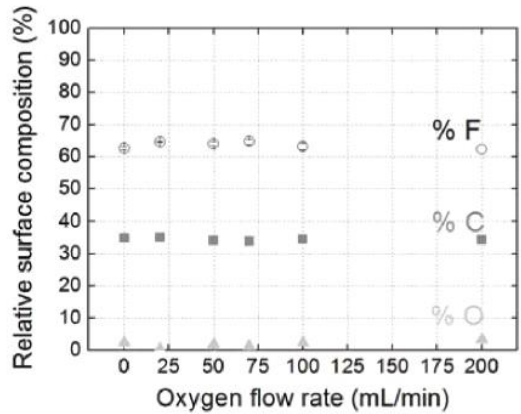 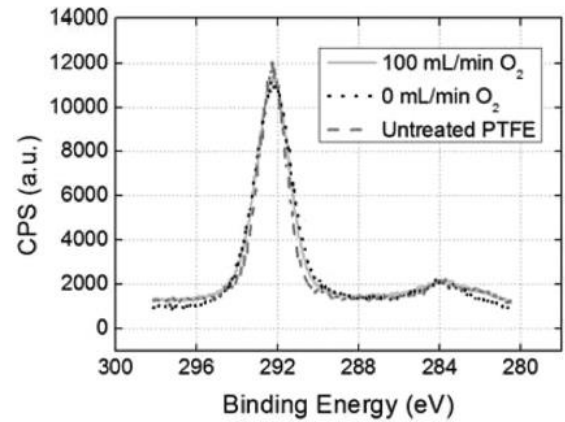

Figure 2. Surface elementary compositions of He and He–O$_2$ plasma-treated PTFE samples ($v_s$ = 25 mm/s, $L_s$ = 10 mm, $g_s$ = 500 µm, $N_s$=1000, $\Phi$(He) = 15 L/min, $P_{RF}$ = 90 W).

Figure 3. HR C 1s spectrum of He and He–O$_2$ plasma-treated PTFE ($v_s$ = 25 mm/s, $L_s$ = 10 mm, $g_s$ = 500 µm, $N_s$ = 1000, $\Phi$(He) = 15 L/min, $P_{RF}$ = 90 W).

As for both treatments (with and without oxygen) there is no modification of the surface composition (as recorded by XPS) but a change in the wetting of the surface, we have decided to investigate these two different phenomena. The He−O$_2$ plasma treatment of PTFE surfaces has already been shown to be responsible for an etching process of the surface.6 However, the effect of pure He plasma is completely different. The observations on XPS (no change) and dWCA (slight decrease) results seem to indicate a smoothing of the PTFE surface with no modification of stoichiometry.

To confirm the existence of this surface modification, AFM images have been recorded, as the surface energy can be linked to the roughness of the sample according to the Wenzel or Cassie−Baxter models.17 As seen in Figure 4, the dWCA results can be correlated to the AFM images. Indeed, almost no differences in the roughness are observed between the untreated sample (a) and the sample treated in pure helium (b); while an increase in the roughness is highlighted in the presence of O$_2$ at a flow rate of 150 mL/min (c). The 2 nm roughness decrease between panels a and b in Figure 4, due to the smoothing of the PTFE surface with a pure helium plasma, is consistent with a decrease in the contact angle observed in Figure 1 and with the theoretical predictions that could be extracted from (for instance) the Wenzel equation. However, this roughness decrease should be taken with great care due to the small absolute value of the changes measured and to the uncertainties of both the AFM and the dWCA measurements.

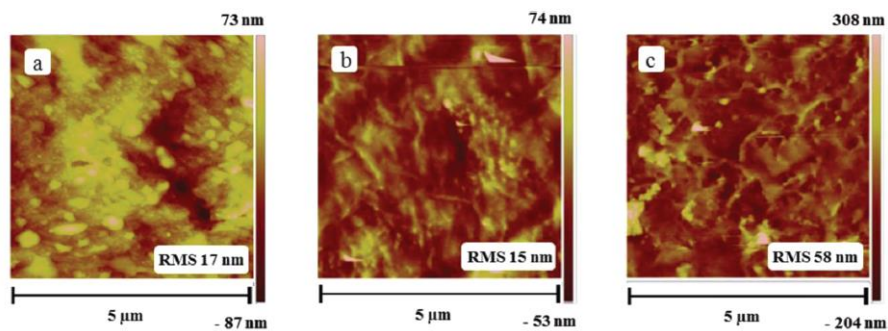

Figure 4. 5×5 µm$^2$ AFM height images of untreated (a) and post-discharge treated PTFE samples with (b) $\Phi$(O$_2$) = 0 mL/min and (c) $\Phi$(O$_2$) =150 mL/min. The experimental conditions are $v_s$ = 25 mm/s, $L_s$ = 10 mm, $g_s$ = 1 mm, Ns =1000, $\Phi$(He) = 15 L/min, $P_{RF}$ = 90 W. The values of the rms roughness are given.

The alveolar structures observed in the presence of oxygen (see Figure 4 and ref 6) could be linked to the crystalline structure of the polymer. Indeed, the species existing in the post-discharge (supposed to be atomic oxygen) preferentially etch the amorphous region.26 This could also be confirmed by the decrease in amorphous PTFE peak observed in FTIR (740 cm$^{-1}$) when $O_2$ molecules are added to a fluorinated monomer used to deposit polymer films.27

## 3.2. Indirect measurements of the surface modifications of PTFE

As the pure helium post-discharge seems to have no significant influence on the chemistry of PTFE surface (regarding dWCA, AFM, and XPS results) and in order to explain this phenomenon, other measurements have been performed. We first studied the relative mass losses of PTFE samples versus the oxygen flow rate. A surprising result is obtained as the highest mass loss is measured for a treatment with a pure helium post-discharge (Figure 5). Indeed, from the literature, an etching with an atmospheric plasma jet of similar technology is attributed to the presence of oxygen atoms (the gas mixture is He containing at least 1% oxygen).28 According to the dWCA and XPS analyses realized on PTFE, this particular behavior was unexpected, as no real modification of the surface properties was observed.

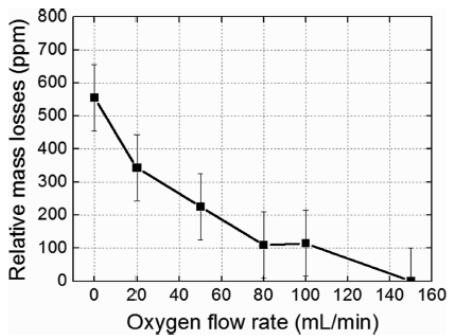
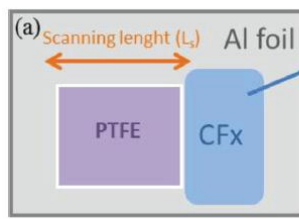
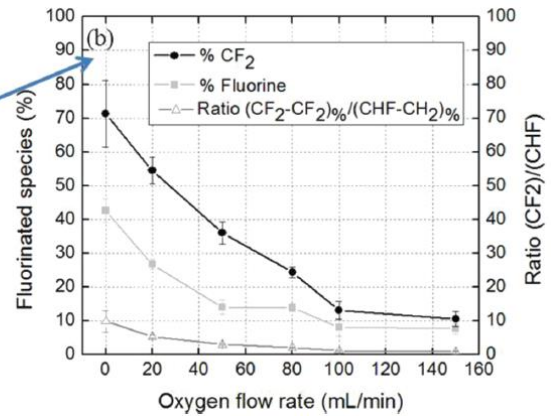

Figure 5. Relative mass losses (ppm) of PTFE samples as a function of the oxygen flow rate ($v_s$ = 25 mm/s, $L_s$ = 10 mm, $g_s$ = 1 mm, $N_s$ =1000, $\Phi$(He) = 15 L/min, $P_{RF}$ = 90 W).

Figure 6. (a) Scheme of the process of detection of fluorinated species on aluminum foil. (b) Relative surface composition of fluorinated fragments on aluminum foil ($v_s$ = 25 mm/s, $L_s$ = 10 mm, $g_s$ = 1 mm, $N_s$ = 1000, $\Phi$(He) = 15 L/min, $P_{RF}$ = 90 W).

In order to complete the mass losses results, we placed aluminum foil (known to be an efficient fluorine trap24) close to the PTFE during the treatment, as illustrated in Figure 6. The surface composition of the Al foil has been characterized at a distance of around 6 mm from the PTFE sample. The fluorine percentage is calculated from the survey spectra, the $CF_2$ amount from the HR C1s spectrum, and the $CF_2-CF_2/CHF-CH_2$ ratio is determined from the areas of the HR F1s.

The relative compositions of fluorine and $CF_2$ fragments versus the oxygen flow rate are in good correlation with mass losses results. Indeed, as higher mass losses (from the PTFE sample) are measured during the treatment without oxygen, more fluorinated compounds are detected on the aluminum foil. In Figure 6, the $CF_2-CF_2/CHF-CH_2$ ratio shows that most of the fluorine compounds are under the form of $CF_2$ when no oxygen is supplied. In order to underline these differences, HR spectra of fluorine and carbon were performed.

The HR F 1s spectrum (Figure 7) is characterized by two components at 689.7 and 686.7 eV, where the first one can easily be attributed to the fluorine from a $-CF_2-CF_2-$ structure (PTFE).[29,30] The appearance of a peak at lower binding energy in the HR F 1s spectrum cannot be associated with a bond between fluorine and aluminum. Indeed, if such compounds as $AlF_3$ (F 1s = 687.75 eV)[31] were formed, the Al 2p peak should have moved to higher binding energy (from 74.68 to 77.27 eV),[31] which is not the case. According to the literature,[29,30] this second peak at 686.7 eV could match with a $-CHF-CH_2-$ structure (polyvinyl fluoride (PVF)) where the main source of hydrogen and $CH_2$ functions are supposed to be the adventitious carbon contamination of the aluminum foil. Moreover, it is important to note that the deconvolution of the fluorine spectrum was operated with two components, but a third component could be added in small amounts. Indeed, the formation of PVDF-like ($-CF_2-CH_2-$) fragments could also be invoked as their binding energy is 688.15 eV, comprised between PTFE and PVF.[29,30]

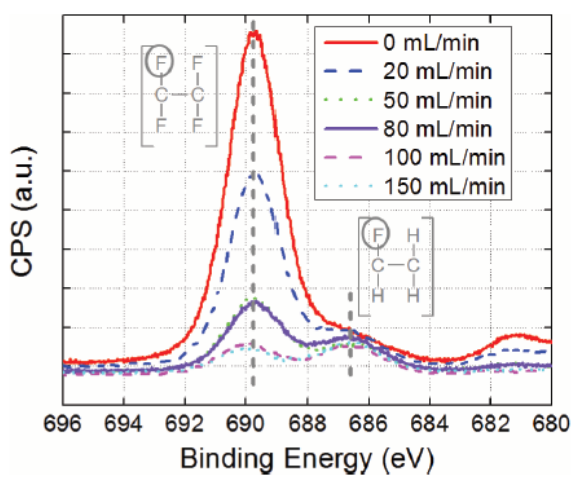

Figure 7. HR F 1s spectrum of aluminum foil placed close to the PTFE sample ($v_s$ = 25 mm/s, $L_s$ = 10 mm, $g_s$ = 1 mm, $N_s$ = 1000, $\Phi(He)$ = 15 L/min, $P_{RF}$ = 90 W).

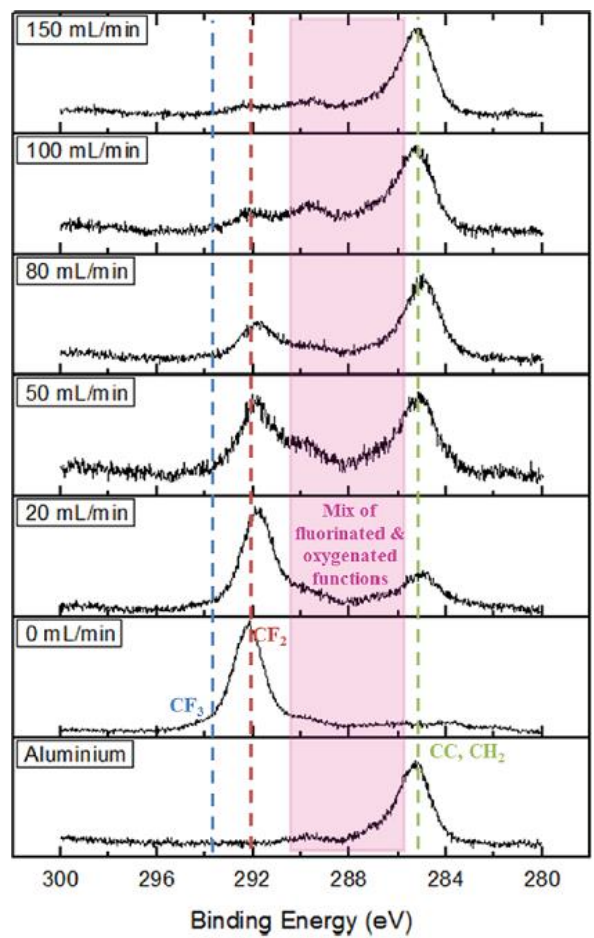

Figure 8. Comparison of the HR C 1s spectra of aluminum foil measured next to the PTFE sample ($v_s$ = 25 mm/s, $L_s$ = 10 mm, $g_s$ = 1 mm, $N_s$ = 1000, $\Phi(He)$ = 15 L/min, $P_{RF}$ = 90 W).

The deconvolution of the HR C1s spectrum (Figure 8) is more complex, as many different components from the PTFE surface (fluorinated species) and from the aluminum foil (oxygenated species) can be observed. The main carbon species that could be present on the surface are listed in Table 1.

| chemical function | C1sBE (eV) | ref |
|---|---|---|
| C–C | 284.6 | 32 |
| C=C | 284.7 | 29 |
| (CH$_2$)$_n$ | 285.0 | 29, 32, 33 |
| CHF–$\underline{C}$H$_2$ | 285.7 | 29 |
| CHF–$\underline{C}$H$_2$ | 285.9 | 34 |
| –(CF$_2$–CH$_2$)$_n$– | 286.4 | 29, 33 |
| C–O–C | 286.45 | 33 |
| C–O or $\underline{C}$–CF | 286.5 | 15 |
| CH$_2$–$\underline{C}$HF–CH$_2$ | 287.7 | 33 |
| C=O or $\underline{C}$HF–CH$_2$ | 287.9 | 29, 33 |
| C=O or CF | 288.0 | 15, 35 |
| $\underline{C}$HF–CH$_2$ | 288.1 | 34 |
| CF | 288.3 | 33 |
| CHF–CHF | 288.4 | 34 |
| O–C=O or HF$\underline{C}$–CF$_2$ or $\underline{C}$F–CF$_2$ | 289.2 | 15 |
| CF | 289.8 | 36 |
| –(CH$_2$–$\underline{C}$F$_2$)$_n$– | 290.9 | 29 |
| –(CHF–$\underline{C}$F$_2$)$_n$– | 291.6 | 33 |
| CF$_2$ or CFO | 292.2 | 37 |
| –(CF$_2$–CF$_2$)$_n$– | 292.5 | 29, 33, 34 |
| CF$_3$ or CF$_2$–O | 294.1 | 27 |
| $\underline{C}$F$_3$–CF$_2$ | 294.6 | 33 |

Table 1. Binding Energy of Various Carbon Species[30]

| O$_2$ flow rate mL/min | surface elementary compositions (atom %) | | | |
|---|---|---|---|---|
| | F | C | O | Al |
| 0 | 43 | 27 | 12 | 18 |
| 20 | 27 | 20 | 26 | 27 |
| 50 | 14 | 12 | 40 | 34 |
| 80 | 13 | 16 | 38 | 33 |
| 100 | 8 | 17 | 39 | 36 |
| 150 | 7 | 20 | 38 | 35 |
| untreated aluminum | 0 | 28 | 38 | 34 |

$^a v_s$ = 25 mm/s, $L_s$ = 10 mm, $g_s$ = 1 mm, $N_s$ = 1000, $\Phi$(He) = 15 L/min, $P_{RF}$ = 90 W. The average standard deviation is ±2%.

Table 2. Surface Elementary Compositions of Aluminum Foil Measured Next to the PTFE Sample

The HR C1s spectrum of the aluminum foil is of great interest for the identification of the ejected fluorinated species from the PTFE surface. The spectrum of the untreated aluminum foil is represented in order to identify the carbon species that do not come from the polymer. Thereby, CF$_2$ fragments can easily be highlighted on the aluminum foil at a binding energy around 292.2 eV. CF$_3$ can also be detected for low oxygen flow rates at 294 eV. The peak around 285 eV can be attributed to hydro-carbonated functions probably coming from the aluminum substrate. The peaks comprised between C–C and CF$_2$ species (i.e., 286–291 eV) are assumed to be a mix of fluorinated and oxygenated functions, as both elements are detected in the spectra (see Table 2). Their binding energies being too close to be resolved, we did not attempt a peak fitting with the 14 possible components identified in Table 1. However, a peak around 290 eV could be attributed to fluorinated species, as it is more noticeable when more fluorine is detected on the aluminum foil. According to Table 1, the most plausible species should be CF at 289.8 eV, or CHF–CH$_2$ at 289.2 eV, as it is observed in the HR F 1s (Figure 7) that fluorine could be linked to a hydro-carbonated backbone.

The surface elementary composition of the untreated aluminum was also measured to compare the coverage of the foil. According to the amount of Al detected on the different samples, we can highlight a more important deposit of polymer with a pure helium plasma, as a lower atomic concentration of aluminum is measured.

One very interesting observation in the HR C1s and F1s spectra is the different behaviors observed regarding the CF$_2$ fragments and the aluminum background (through the C–C component). The increase in oxygen flow rate induces a lower deposition of fragments from PTFE. Indeed, even if 7% fluorine and 9% CF$_2$ are detected at the surface of aluminum for Φ(O$_2$) = 150 mL/min, the profile of C1s is close to the one from the untreated aluminum foil. When no oxygen is injected in the plasma, the peaks associated with the presence of the aluminum foils (C–C, oxygenated carbon) are almost included in the background noise, and CF$_2$ is the main component, as it represents 70% of the C1s peak. Moreover, the ejection of a higher amount of material in the pure He plasma can easily be highlighted by the attenuation of the signal coming from the aluminum substrate (Table 2). Regarding the C1s and F1s peaks and mass losses, we suggest that the species existing in the He plasma are more energetic as higher fragmentation of PTFE is performed. However, these species do not seem to have a chemical effect, as no surface modification of the PTFE is observed. Concerning the ratio of CF$_2$–CF$_2$/CHF–CH$_2$, it is not surprising to detect more CF$_2$ than CF because C–C bonds are more easily broken than C–F (3.8 and 4.7 eV bond energies).[38]

### 3.3. Species involved in the PTFE surface modification

From the two different kinds of treatments operated (with and without oxygen), we propose to highlight the role of the species involved in the etching process operating at the post-discharge/PTFE interface (Figure 9). The comparison of the RF He and He-$O_2$ post-discharges seems to be an interesting approach to describe them. The characterization and the evolution of different species as a function of oxygen flow has previously been studied.39 Considering the evolution of dWCA, the mass losses (correlated with fluorinated species detected on the aluminum foil), and the species in the afterglow, an explanation of these behaviors can be advanced.

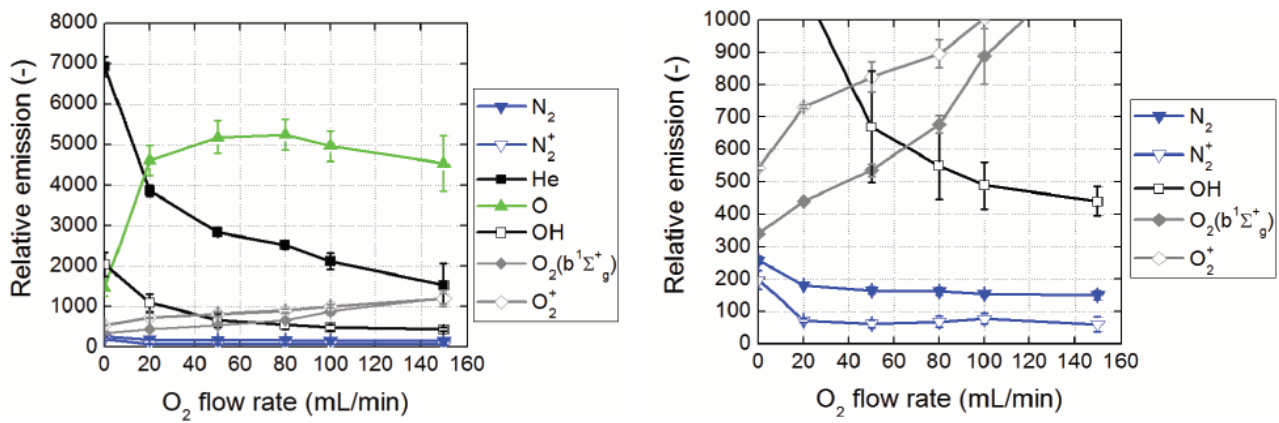

Figure 9. (a) Optical emission spectra of the whole post-discharge versus the $O_2$ flow rate in the presence of PTFE. (b) Zoom of panel a. ($\Phi$(He) = 15 L/min, $P_{RF}$ = 120 W, $g_s$ = 1 mm).

By considering all the species detected in the post-discharge during the treatment of the PTFE surface, we focused our work on the ones most probably responsible for the behaviors observed. Indeed, nitrogen species, oxygen ions, and OH molecules are known to involve totally different surface modifications as nitrogenated3,8,40,41 or oxygenated8,42 functions are created in the PTFE, and the WCA is decreased.48,49 Moreover, the emission intensities (although not directly related to the concentration) of some of these species are so low that no significant effect is expected from them. The considered species in this analysis are the atomic oxygen, oxygen metastables, and helium metastables.

**Atomic Oxygen**

According to the literature9,43–46 in oxygen plasmas, oxygen atoms seem to be the main active species in the etching of organic films. Wydeven et al.46 studied the O($^3$P) etching behavior in PTFE and showed by XPS measurements that the polymer exhibited only minor surface oxidation and a very slight decrease in F/C ratio from 2.00 to 1.97. The assumption of oxygen atoms being the etching species seems to be reliable, as these results are consistent with our experiments.

**Oxygen Metastables**

It has been shown45 that the Σ metastable oxygen molecule $O_2(b^1\Sigma^+_g)$ should be unreactive toward organic compounds, and converted into the Δ state $O_2(a^1\Delta_g)$ by collision with other molecules. In this study dealing with the etching of polyimide, the evolution of these last species was not consistent with the etch rate of the films. The Δ state emitting between 1.2 and 1.3 µm43 cannot be evidenced by conventional spectrometers and thus cannot be confirmed for our PTFE samples. However, we suppose that $O_2(b^1\Sigma^+_g)$ and $O_2(a^1\Delta_g)$ metastables cannot be responsible for the physical etching, as their energy levels are low (1.63 and 0.98 eV, respectively)47 compared to the C–C and C–F bond energies (3.8 and 4.7 eV).38

**Helium Metastables**

The treatment of PTFE with the plasma torch supplied in helium without oxygen generates an (important) mass loss. This is not surprising, as it has been shown that helium can be responsible for the sputtering of PTFE.21,22 Indeed, these authors measured fluorocarbon species present in the gas phase with a UV emission spectrometer and analyzed the fluorinated fragments deposited onto the substrates by means of XPS and infrared spectroscopy. However these researchers were only interested in the deposition of PTFE-like from a sputtering of a PTFE target, and they did not analyze the target surface. Another study50 highlighted the ejection of solid polymer and mass loss when a PTFE sample was submitted to a glow discharge supplied in helium. We are not able to compare the characterization of the treated surface, but other studies appear to be correlated with our behavior.50,51 Indeed, on one hand, the TFE monomer ($CF_2=CF_2$) seems to be the main product in the helium plasma-induced decomposition of PTFE.50 On the other hand, the RF plasma sputtering of PTFE is considered to involve scission of $-(CF_2)_n-$ chains yielding smaller $-(CF_2)_m-$ ($m \ll n$) segments, which are deposited at the substrate surfaces and form a structure dominated by $-(CF_2)-$ groups.51 The species that we assume being responsible for this physical etching are the high energetic helium metastables. Although their presence in a pure He discharge has been known for a long time, they were experimentally evidenced in our setup in a recent study,39 and they can also be emphasized by the observation of $N_2^+$. Indeed, it has been mentioned that $N_2$ molecules easily quench the $He^m$ to create $N_2^+$ through the Penning ionization of $N_2$ 52,53

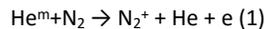
$$He^m + N_2 \rightarrow N_2^+ + He + e \quad (1)$$

According to our results in a pure He plasma, since no modification of the chemistry (Figures 2 and 3) and no real change of wettability (Figure 1) and roughness (Figure 4) are observed in spite of mass losses (Figure 5), we assume a layer-by-layer physical etching with no chemical reactions. As the capability of helium to sputter the PTFE has been highlighted and as it is known to be an efficient gas to cross-link a polymer,54,55 the assumption of the He metastables seems quite consistent. Helium is an inert gas having mainly structural effects as degradation by scissions or cross-linking56 but should not react chemically with the surface of polymer. Helium metastable species being highly energetic (19.82 and 20.62 eV for $2^3S$ and $2^1S$ helium states, respectively),57 they can easily break organic bonds. Energies of activated metastables are much larger than the C–C and C–F bond energies (3.8 and 4.7 eV).38

According to the XPS analyses of PTFE surfaces (Figures 2 and 3), the treatment in pure helium preferentially induces a scission of C–C bonds over C–F. It is not surprising that this behavior is completely different from other treatments such as electron beams, X-ray, or UV irradiations. Indeed, in these cases, the species responsible for the cleavage of the C–F bond are electrons or secondary electrons produced by the interaction of the radiation with the polymer,58 because of the high electron affinity of the fluorine. We suggest that the helium metastables being neutral, they do not have a special affinity for the C–F bonds and should statistically preferentially break the more fragile C–C bond. In the transfer of the potential energy of the He metastables to the total molecule, the much more fragile C–C bonds are therefore expected to absorb the released energy first. More theoretical and experimental investigation will, however, be required in the future to fully understand the lack of reactivity of the C–F bonds, that should also be, at least, partly attacked.

Contrary to treatments in pure helium, the addition of oxygen into the discharge implies wettability and roughness modifications generated by an anisotropic etching of PTFE surfaces. It is well-known that the addition of an oxidant gas (e.g., oxygen) usually leads to a hydrophilic surface as polar functional groups are introduced into the polymer surface.59,60 In the case of PTFE, the behavior is more controversial as $O_2$-containing plasmas can lead to more hydrophobic[6–10] or hydrophilic surfaces.[11,12] However, the PTFE treated samples showing an increase in hydrophobicity in the literature seem to have the same behavior regarding chemistry and morphology as our samples. No modification has been highlighted by XPS analyses, and an increase in roughness has been observed. As oxygen atoms are able to etch organic films, we assume that they are responsible for the (physical and/or chemical) etching of PTFE surfaces. Contrary to helium, $O_2$ is a reactive gas involving species (atomic oxygen and oxygen metastables) that can react chemically with the polymer creating surface modifications. Moreover, as already mentioned, oxygen atoms etch preferentially the amorphous phase and consequently lead to a roughening of the surface. In order to have a better understanding and visualization, the two processes have been depicted in a schematic representation (Figure 10).

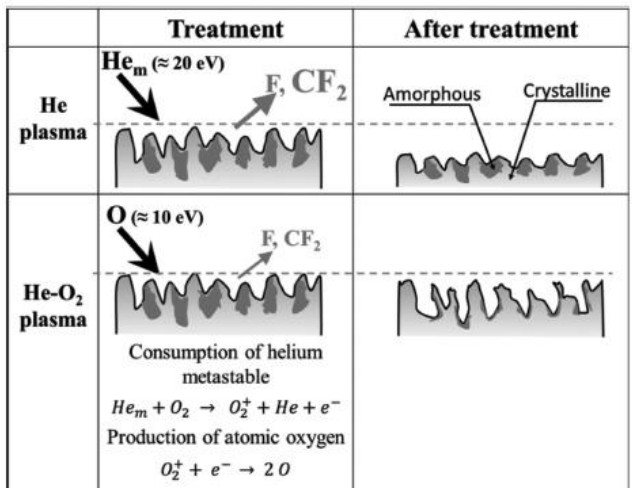

*Figure 10. Scheme of the two possible etching processes. In a pure helium plasma, the most important fragment produced is $CF_2$.*

Concerning the decrease in species ejected versus the $O_2$ flow rate (Figures 5–8), we suppose that it is due to the decrease in helium species. Indeed, Léveillé et al.[61] have supposed that the addition of an electronegative gas (e.g. $O_2$) causes a reduction of the electron density, which can no longer participate to the excitation of He radiative states. Moreover, we previously showed[39] that the concentration in helium metastable species decreases versus the oxygen flow rate as they are consumed through the Penning ionization of $O_2$.[62] This decay is also in agreement with a previous study[63] where the density of the $He(2^3S)$ metastable as a function of the oxygen concentration was measured in a He–$O_2$ microwave discharge.

$$He^m + O_2 \rightarrow O_2^+ + He + e \quad (2)$$

Fewer helium metastable species are then available to etch the PTFE, generating a reduction in mass losses for higher $O_2$ flow rates. Therefore, it seems consistent that the He$^m$ involve larger mass losses than O atoms as the Gotrian diagrams of He I and O I indicate that radiative states of helium present energetic levels almost twice higher[64] (the upper energy level of O at 777 nm being 10.74 eV).[65]

## 4. Conclusion

The different behavior of a PTFE surface exposed to the afterglow of a He and He−$O_2$ RF plasma torch working at atmospheric pressure has been studied by means of three ways. In He−$O_2$ plasmas, no modification of the PTFE surface was highlighted by XPS analyses, but an increase in dWCA and in roughness was observed. In this case, we assume an anisotropic etching where the oxygen atoms mainly etch the amorphous phase. In pure He plasma, as no modification of the chemistry and no significant change of wettability and roughness were highlighted in spite of important mass losses, we assume a layer-by-layer physical etching without any preferential orientation. The highly energetic helium metastables are supposed to be the main species responsible for the scission of –$(CF_2)_n$– chains. Indeed, the helium metastable species being consumed by reaction with oxygen, fewer of them are then available to etch the PTFE when more $O_2$ is supplied. This has been proved by the measurements of mass losses and the detection of fluorinated species onto aluminum foil as less species are ejected from the PTFE when more $O_2$ is injected into the plasma.

## 5. Acknowledgments

This work was part of the I.A.P (Interuniversitary Attraction Pole) program "PSI – Physical Chemistry of Plasma Surface Interactions" financially supported by the Belgian Federal Office for Science Policy (BELSPO). This work was also financially supported by the FNRS (Belgian National Fund for Scientific Research) for an "Aspirant Grant (FRFC Grant No. 2.4543.04)", Région Wallonne (OPTI2MAT Excellence Program), and the European Commission (FEDER – Revêtements Fonctionnels).